\documentclass[%
 reprint,
 amsmath,amssymb,
 aps,
]{revtex4-1}

\usepackage{graphicx}% Include figure files
\usepackage{dcolumn}% Align table columns on decimal point
\usepackage{bm}% bold math
\usepackage{gensymb}
\usepackage{siunitx}
\usepackage[caption=false]{subfig}
\usepackage{hyperref}% add hypertext capabilities
\begin{document}

\preprint{APS/123-QED}

\title{Juggling with light}% Force line breaks with \\

\author{Albert J. Bae}
\altaffiliation[Present address: ]{%
    Department of Biomedical Engineering,
    University of Rochester,
    Rochester NY 14627,
    USA.
}%Lines break automatically or can be forced with \\
\affiliation{%
    Max Planck Institute for Dynamics and Self-Organization,
    37077 Goettingen,
    Germany.
}
\author{Dag Hanstorp}
\author{Kelken Chang}
\email{Corresponding author. kelken.chang@physics.gu.se}
\affiliation{%
    Department of Physics,
    University of Gothenburg,
    412 96 Gothenburg,
    Sweden.
}%

\date{\today}% It is always \today, today,

\begin{abstract}
We discovered that when a pair of small particles is optically levitated, the particles execute a dance whose motion resembles the orbits of balls being juggled. This motion lies in a plane perpendicular to the polarization of the incident light. We ascribe the dance to a mechanism by which the dominant force on each particle cyclically alternates between radiation pressure and gravity as each particle takes turns eclipsing the other. We explain the plane of motion by considering the anisotropic scattering of polarized light at a curved interface.
%\begin{description}
%\end{description}
\end{abstract}

\pacs{Valid PACS appear here}% PACS, the Physics and Astronomy
%Use showkeys class option if keyword
\maketitle

\begin{figure}[htbp]
% trim the figure {left bottom right top}
\includegraphics[width= 0.7 \columnwidth, trim={37 67 8 35}, clip]{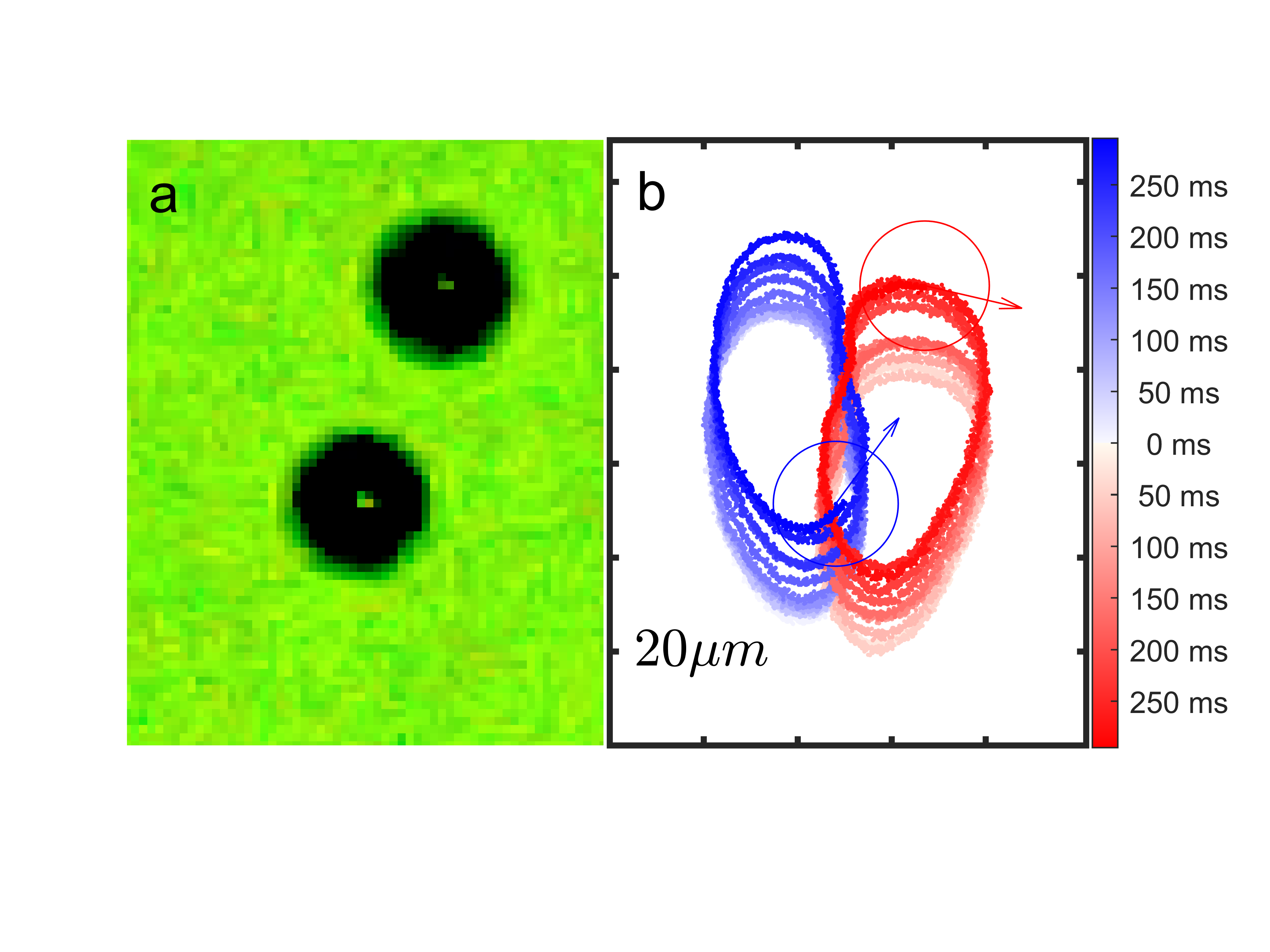}
\caption{\label{fig:JugglingDroplets} (a) An image of a pair of optically levitated glycerol droplets and (b) their trajectories. The beam polarization vector is perpendicular to the page. Arrows indicate the instantaneous velocity at the snapshot of time corresponding to panel (a). The color bar indicates time in milliseconds.}
\end{figure}

\begin{figure}[htbp]
% trim the figure {left bottom right top}
\includegraphics[height= 0.35 \columnwidth]{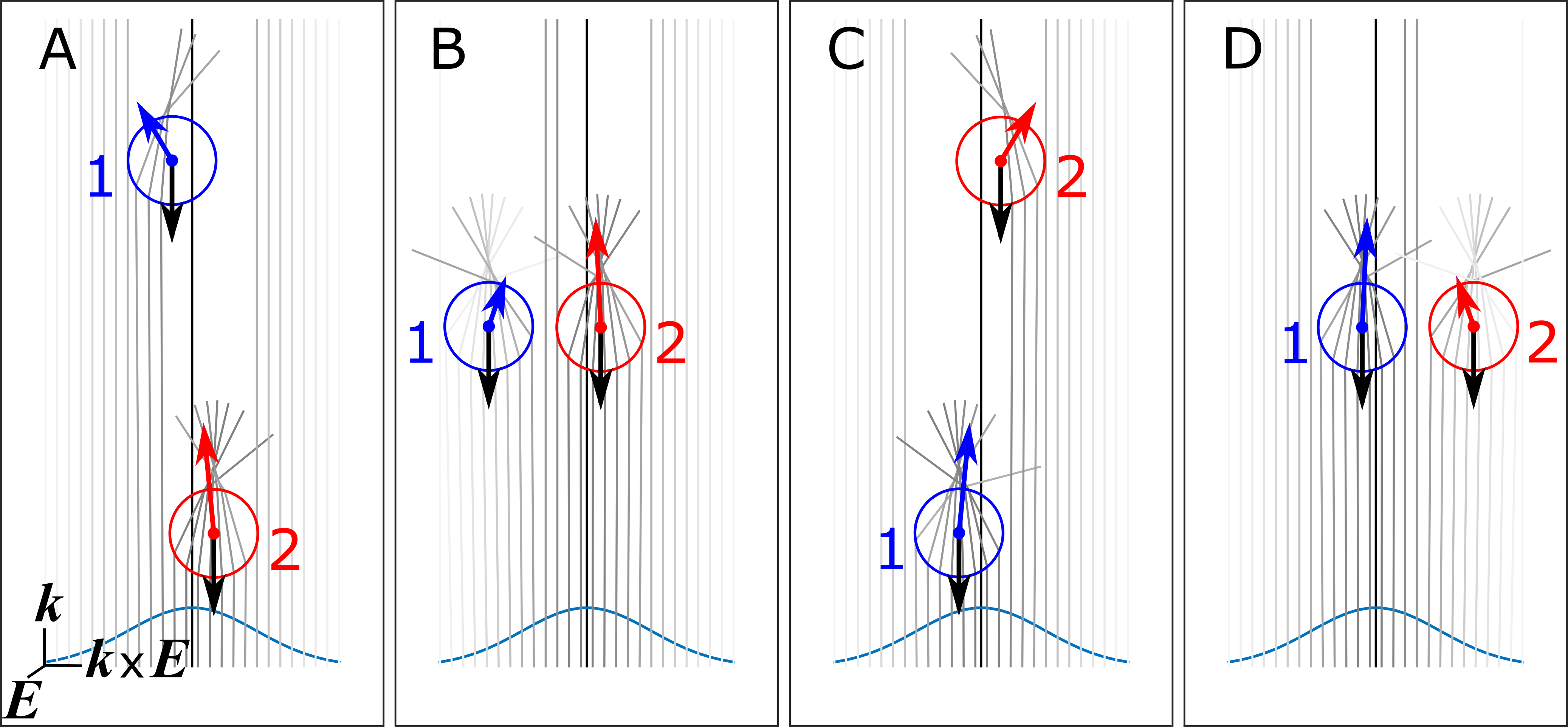}
\caption{\label{fig:JugglingCycle} Two particles juggle in a beam of light propagating with wave vector $\boldsymbol{k}$ and polarization vector $\boldsymbol{E}$. Black arrows denote gravity, while red or blue arrows denote the optical forces. The blue Gaussian curve at the bottom denotes the beam intensity profile. (A) Particle {\sf 1} is eclipsed by {\sf 2}, it experiences less light and falls, while the unobstructed particle {\sf 2} receives more light than its weight and rises. (B) Particle {\sf 1} is no longer eclipsed by {\sf 2}, the gradient force pushes it back towards the centerline. Particle {\sf 2} continues to move upward. In panels (C) and (D), the roles of particle {\sf 1} and {\sf 2} reverse.}
\end{figure}

\begin{figure}[htbp]
% trim the figure {left bottom right top}
\subfloat{
\includegraphics[width= 0.4 \columnwidth, trim={0 -70 0 0}, clip]{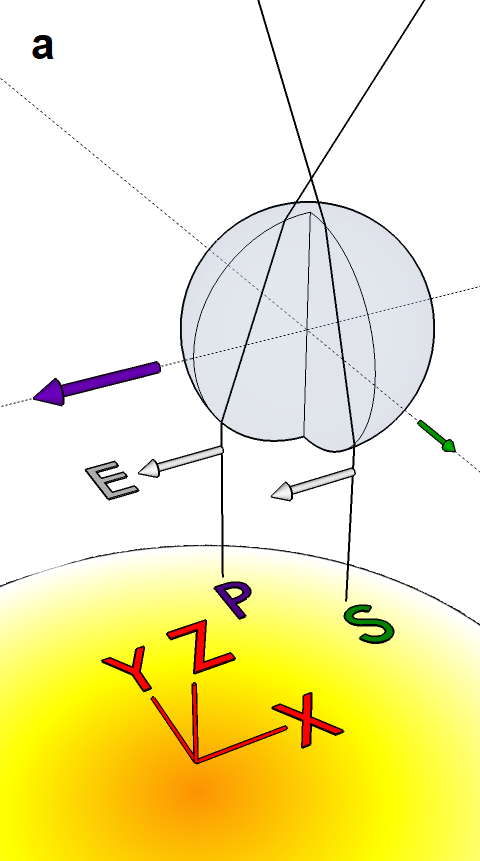}
}
%\hspace{0.5pt}
\begin{minipage}[b]{0.5 \columnwidth}
\raggedright
\subfloat{
\includegraphics[width= 0.87 \columnwidth, trim={47 0 50 15}, clip]{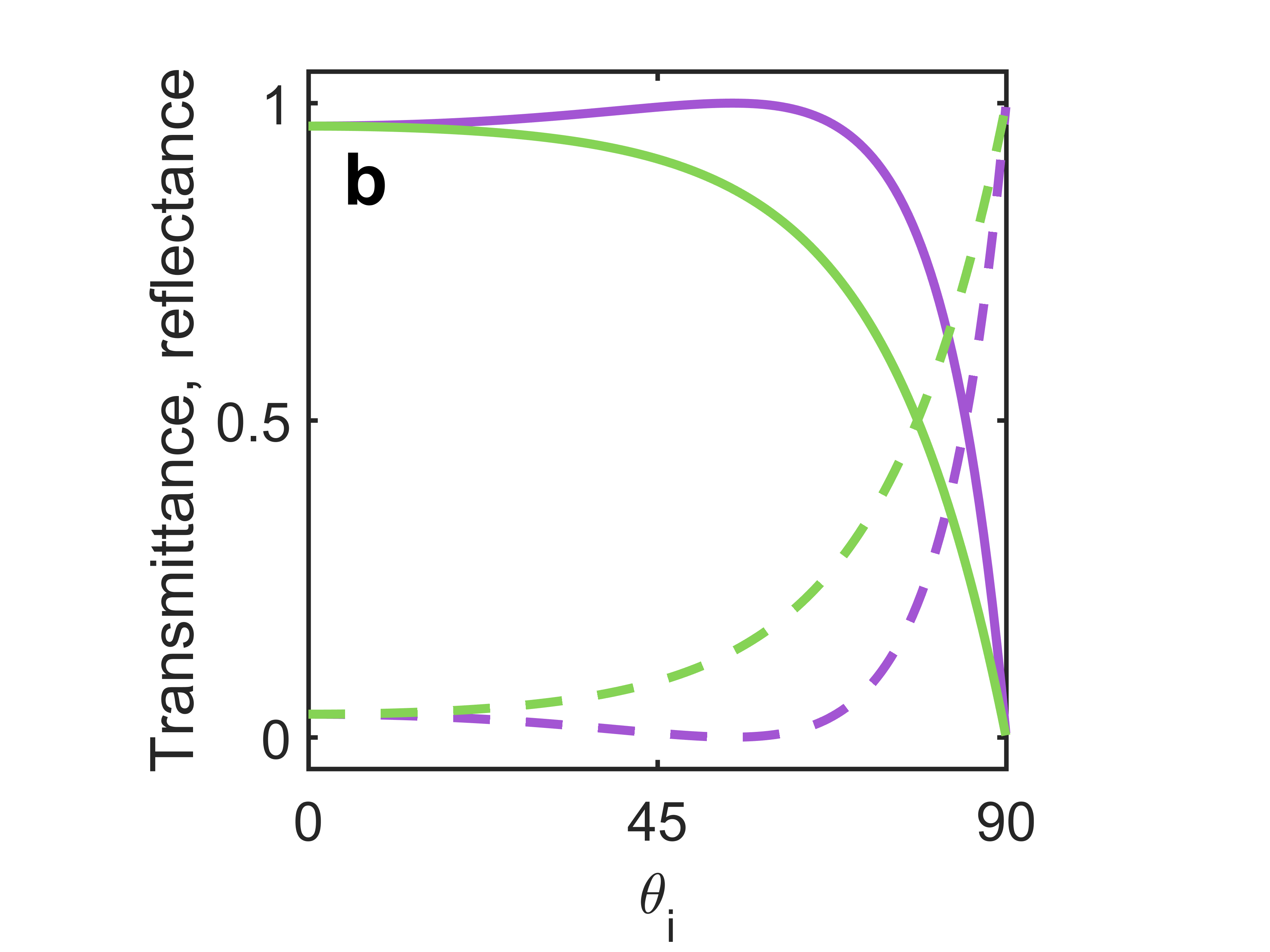}
}\\[2pt]
\subfloat{
\includegraphics[width= \columnwidth, trim={12 -8 24 15}, clip]{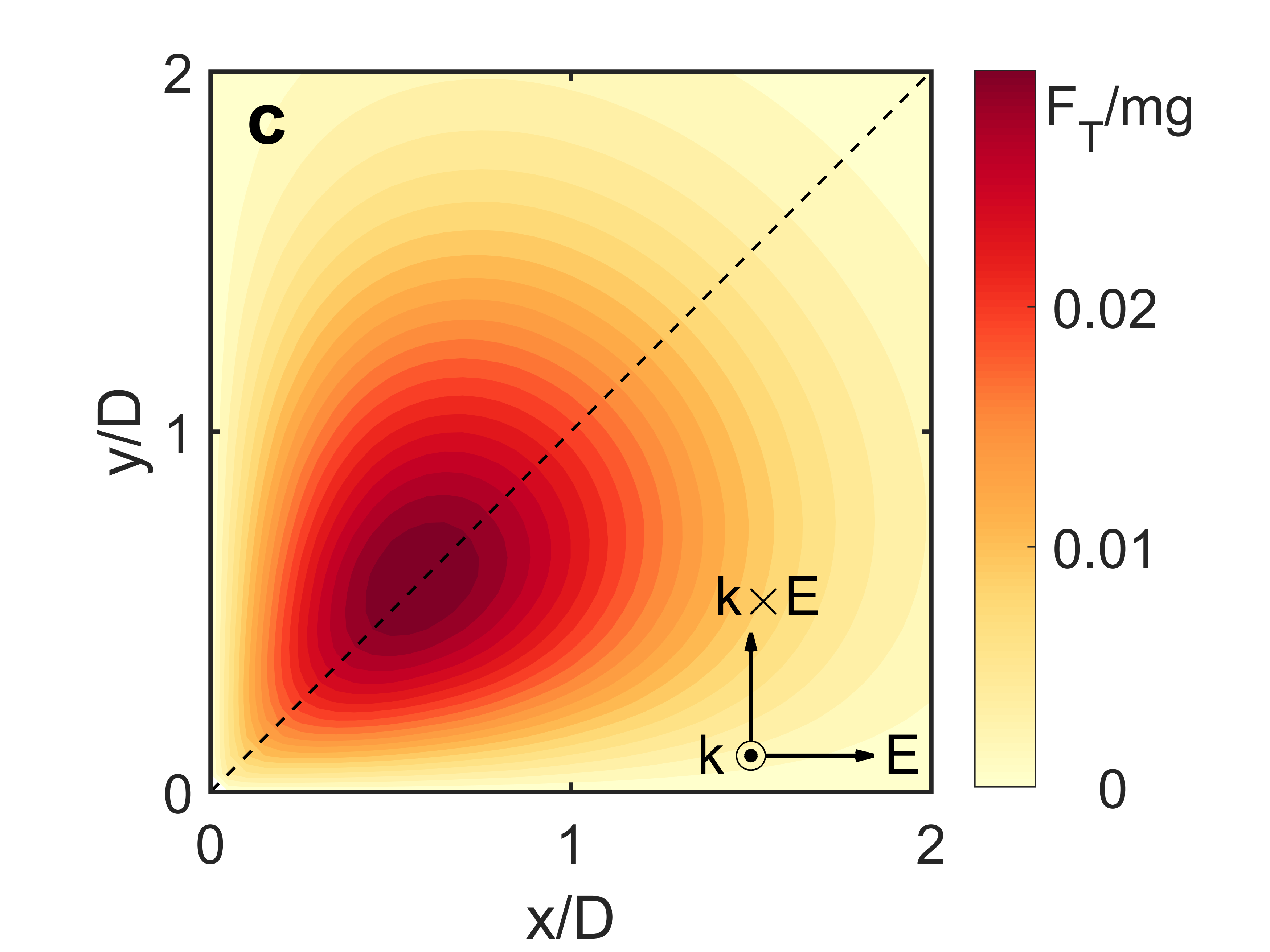}
}
\end{minipage}
\caption{\label{fig:PolarizationForces}(a) Two representative equal-power, linearly polarized rays enter the sphere with different orientation of the polarization vector $\boldsymbol{E}$. (b) Transmittances (solid lines) and reflectances (dashed lines) for $p$-polarized (purple lines) and $s$-polarized light (green lines) at an air-glycerol interface as a function of angle of incidence $\theta_{\rm i}$. (c) Ray tracing calculation of the tangential forces ($F_T$), in units of the gravitational force $m g$, for a \SI{28}{\micro\metre} glycerol droplet as a function of the droplet position $\boldsymbol{r}$. The incident beam wave vector $\boldsymbol{k}$ points out of the page. All forces point in the counter-clockwise direction. They are strongest at a $45\degree$ polar angle (dashed line).}
\end{figure}

The idea of using light to propel particles has been a topic of study as early as the 17$^\mathrm{th}$ century when Johannes Kepler hypothesized that solar radiation was responsible for pushing the comet's tail away from the sun. Light propulsion regained great relevance in the 1970s when Arthur Ashkin discovered the optical tweezers. In a series of seminal articles, Ashkin laid down the experimental, conceptual and theoretical framework for his discovery \cite{Ashkin1970, Ashkin1971, Ashkin1974, Ashkin1975, Ashkin1986, Ashkin1992}. Yet one of his notable observations is barely known. In an article published over forty years ago \cite{Ashkin1975}, he noticed that a levitating laser beam can propel a pair of droplets equal in size to come side by side and briefly touch before they coalesce. Unable to further pursue this research with the technology of his time, Ashkin urged researchers to resolve the puzzle of the colliding droplets with high-speed photography. Recently, there has been renewed interest in this attempt \cite{CarmonaSosa2016, Moore2016, Mitra2018}. Notably, Moore \textit{et al.} \cite{Moore2016} have observed oscillations of two silica particles for up to a few minutes. We were able to finally achieve the demanding spatial and temporal resolutions necessary to observe the droplet motion by constructing an optical levitation setup that includes a long-distance microscope and high-speed movie camera. To our amazement, we discovered that instead of colliding directly, pairs of droplets will frequently execute a dance \footnote{See Supplemental Material at \url{https://youtu.be/ZyXBM8B0Md0} for a movie of the droplets juggling in the laser beam. It has come to our attention that Moore \textit{et al.} observed an oscillations under quite different conditions. Our theory not only explains their observation, but also makes predictions on their particle charge and oscillation frequency.} which may last for up to half an hour during which the droplets move in well-defined planar orbits (Fig.~\ref{fig:JugglingDroplets}).
We call this optical juggling as the motion resembles the orbits of balls being juggled by a carnival performer \cite{Polster2003}. What is responsible for these intricate movements, and what determines the plane in which they lie? 
%\begin{video}
%\href{https://youtu.be/ZyXBM8B0Md0}{\includegraphics[width= \columnwidth, trim={37 67 8 35}, clip]{trajectories_sm_power_1100mW_pulsewidth_100microsec_voltage_60V_06_ReplayRate_200Hz_ver_00.png}}
% \setfloatlink{https://youtu.be/ZyXBM8B0Md0}%
% \caption{\label{vid:JugglingDroplets}%
%  The panel on the left shows an image of a pair of optically levitated glycerol droplets. The panel on the right shows trajectories of the juggling droplets. Arrows indicate the instantaneous velocity at the snapshot of time corresponding to the panel on the left. The color bar indicates time in milliseconds.
% }%
%\end{video}
\\
\indent
Our experiment (Fig.~\ref{fig:Schematic} and \footnote{See Supplemental Material at [URL] for additional details, which includes Refs.~\cite{Atherton1999, Mordant2004, Shampine1997}.}) is similar to Ashkin's original experiment \cite{Ashkin1971, Ashkin1974, Ashkin1975}.
\begin{figure}[htbp]
% trim the figure {left bottom right top}
\includegraphics[width= 0.48\columnwidth]{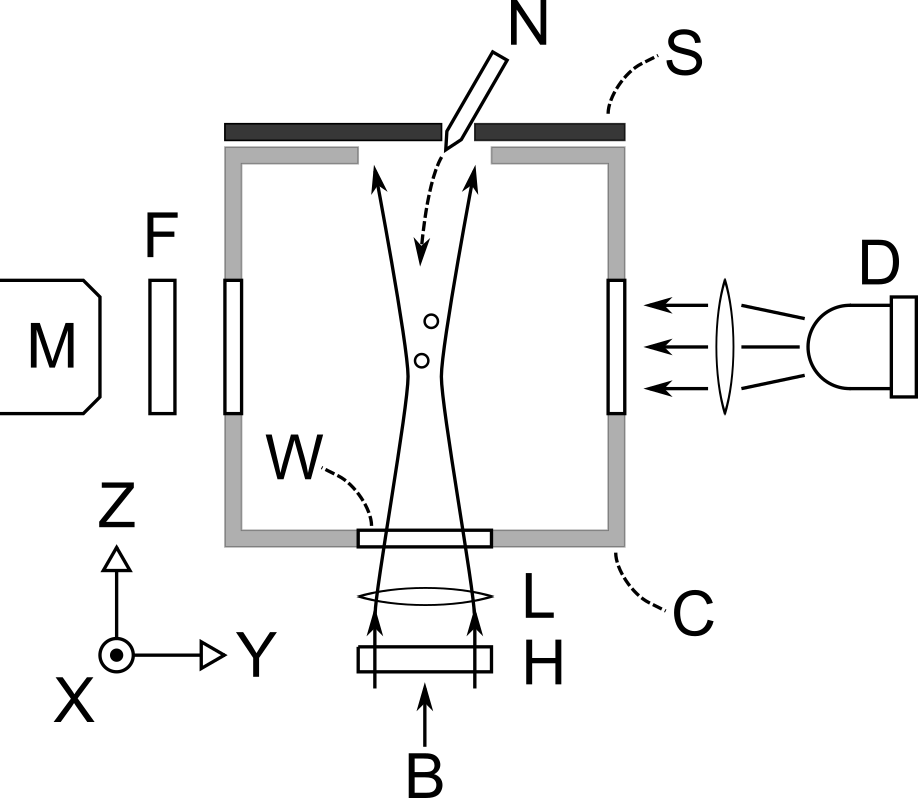}
\caption{\label{fig:Schematic} The labels denote laser beam (B), aluminum chamber (C), LED light source (D), notch filter (F), half-wave plate (H), focusing lens (L), microscope (M), piezoelectric nozzle (N), aluminum sheet (S) and windows (W). The beam polarization is set to be in the X direction.}
\end{figure}
We use a lens L of \SI{5.0}{\centi\metre} focal length to focus \SI{1.0}{\watt} of a \SI{532}{\nano\metre} continuous wave vertical laser beam B. The beam intensity profile is Gaussian and its initial diameter is $0.85 \pm$\SI{0.1}{mm}. The beam is linearly polarized with a polarization direction that can be rotated in the XY plane using a half-wave plate H. Unless we state otherwise, the beam polarization is set to be in the X direction. The beam enters an air-filled aluminum chamber C through a bottom window W. A piezo-electrically controlled nozzle N sequentially produces droplets from a mixture containing 90\% by volume of distilled water and 10\% of glycerol. The nozzle produces droplets that are naturally charged, through a process known as electrostatic spraying \cite{Bailey1988}. The tip of the nozzle is inserted into the chamber through a \SI{5}{\milli\metre}-diameter small hole on an aluminum sheet S covering the top of the chamber. The droplets settle slowly by gravity into the laser beam where they are levitated. Water evaporates during the decent, and by the time the droplet is captured by the beam, the droplet is mostly glycerol with a steady diameter of $28.6 \pm$\SI{2.1}{\micro\metre}. An LED light source D illuminates the droplets from the side and casts the shadow of the droplets into the collecting lens of a long-distance microscope M. A notch filter F in front of the microscope blocks light scattered from the levitating laser beam. A high-speed movie camera is arranged at a right angle to the beam polarization to capture the motion of the droplets in the XZ plane at a frame rate of 45,000~Hz and a spatial resolution of \SI{1.72}{\micro\metre} per pixel. A second set of LED, microscope and camera is arranged at a right angle to the first to synchronously capture the droplet motion in the YZ plane. In total, we captured 61 juggling events. \\
\indent
The droplets dance in beautiful patterns (Fig.~\ref{fig:JugglingDroplets}). With a head-on collision, the droplets begin by leapfrogging down the beam, before eventually overcoming their surface energy barrier to coalesce \footnote{See Supplemental Material at \url{https://youtu.be/UiqqxNjS_v8} for a movie of the droplets coalescing in the laser beam.}. The combined droplet is levitated at a new height below the center of mass of the two initial droplets. With a grazing collision, the droplets also begin with leapfrogging, but they gradually converge to juggling at a stable elevation \footnote{See Supplemental Material at \url{https://youtu.be/uZPhUhgpxxk} for a movie of the formation of juggling movement.}. When the droplets are juggled, they eventually settle in orbits in the same vertical plane that contains the beam axis and perpendicular to the initial polarization vector. The droplets move in pea-shaped orbits. Each orbit measures approximately one droplet diameter in width and two diameters in height. When the droplets come side by side, their separation is about one diameter. We measured the droplet trajectories and found that the droplets experience velocities as great as 40\% of their terminal velocity and accelerations as high as $0.3 g$. We analyzed the droplet velocity spectra and obtained an orbital frequency of $33.9~\pm$\SI{2.7}{\hertz}, in good agreement with a period of $27.7~\pm$\SI{1.3}{\milli\second} obtained from the velocity autocorrelations. Since liquid droplets in air experience little random thermal fluctuations, the droplets are able to juggle for as long as 30 minutes, in excess of 60,000 repetitions prior to coalescence. \\
\indent
How light juggles matter is summarized by the principle: light directs the flow of matter; matter directs the bending of light \footnote{We allude to the elegant quote by John A. Wheeler, ``Spacetime tells matter how to move; matter tells spacetime how to curve''}. Consider two dielectric spheres of mass $m$ subject to the gravitational force $m \boldsymbol{g}$ and a short-range repulsive interaction (Fig.~\ref{fig:JugglingCycle}).
A beam of light with Gaussian intensity profile illuminates the particles. The incident light is linearly polarized along the vector $\boldsymbol{E}$ and propagates upward with wave vector $\boldsymbol{k}$. Initially, particle {\sf 1} is centered slightly to the left of the centerline and particle {\sf 2} slightly to the right, with {\sf 1} above {\sf 2} (Fig.~\ref{fig:JugglingCycle}A). In this configuration, {\sf 1} is eclipsed by {\sf 2}. This obstruction prevents {\sf 1} from receiving sufficient light to overcome gravity, and it falls. The left-hand side of {\sf 1}, which is further from the centerline, receives more light, so the gradient force \cite{Ashkin1986} pushes the particle further away from centerline. Consequently, {\sf 1} moves down and to the left. Since {\sf 2} is close to the centerline, the upward optical force outweighs the downward gravitational force, so {\sf 2} moves upward. Particle {\sf 1} moves out of the shadow of {\sf 2} (Fig.~\ref{fig:JugglingCycle}B), the right-hand side, which is closer to centerline, receives more light, and the gradient force now pushes {\sf 1} back towards the centerline. Particle {\sf 2} is still close to the centerline and continues to move upward. This leads us to Fig.~\ref{fig:JugglingCycle}C, where {\sf 2} has risen above {\sf 1}. Particle {\sf 1} has returned to the centerline, where it casts its shadow on {\sf 2}. The particle positions in parts~\ref{fig:JugglingCycle}A and \ref{fig:JugglingCycle}B are now mirrored in parts~\ref{fig:JugglingCycle}C and \ref{fig:JugglingCycle}D by interchanging {\sf 1} with {\sf 2}, and left with right. After Fig.~\ref{fig:JugglingCycle}D, we end up back to the configuration shown in Fig.~\ref{fig:JugglingCycle}A. This process repeats indefinitely and resembles the motion of balls being tossed in a fountain pattern by a carnival juggler \cite{Polster2003}. \\
\indent
Let us develop this physical picture into a more quantitative description. Consider two glycerol droplets of diameter $D=$~\SI{28}{\micro\metre}, density $\rho =$~\SI{1.26e3}{kg.m^{-3}}, index of refraction $n_2 = 1.47$ and charge $Q=$~\SI{-1.6e-15}{\coulomb} \cite{Note2} immersed in a dielectric medium (air) of permittivity $\epsilon=$~\SI{8.9e-12}{F.m^{-1}}, gravity $g=$~\SI{9.81}{m.s^{-2}}, dynamic viscosity $\eta=$~\SI{1.85e-5}{Pa.s} and index of refraction $n_1 = 1$. The optical forces, gravity, hydrodynamic and electrostatic interactions act together to choreograph the dance. Characteristic values for the forces indicate that optical forces and gravity outweigh both hydrodynamic and electrostatic forces. Clearly, the dominant force must cyclically alternate between optical forces and gravity. Using this information, we constructed the following model. The two droplets obey Newtonian mechanics:
\begin{equation}
\label{eq:eom}
m \frac{{\rm d} \boldsymbol{v}^{(i)}}{{\rm d}t} = \boldsymbol{F}_{G}^{(i)} + \boldsymbol{F}_{H}^{(i)} + \boldsymbol{F}_{Q}^{(i)} + \boldsymbol{F}_{O}^{(i)} \;\; , \;\; (i = 1, 2) \,.
\end{equation}
The gravitational force is given by $\boldsymbol{F}_{G}^{(i)} = - \hat{\boldsymbol{k}} \, \pi \rho g D^3 / 6$.
The hydrodynamic force is given by Stokes' drag containing the lowest order rigid-sphere interaction term \cite{Briels1994}:
\begin{equation}
\boldsymbol{F}_{H}^{(i)} = 3 \pi \eta D \bigg[ -\boldsymbol{v}^{(i)} + \sum_{j \neq i}^{2} \frac{3 D}{8 r_{ij}} \bigg( \boldsymbol{I} + \frac{\boldsymbol{r}_{ij} \boldsymbol{r}_{ij}}{r_{ij}^{2}} \bigg) \cdot \boldsymbol{v}^{(j)} \bigg] \,.
\end{equation}
Here $\boldsymbol{r}_{ij} = \boldsymbol{r}_i - \boldsymbol{r}_j$ is the separation vector between the two droplets, $\boldsymbol{v}$ is the droplet velocity and $\boldsymbol{I}$ is the $3 \times 3$ identity matrix.
The electrostatic interaction between the two identically charged droplets is given by Coulomb's law:
\begin{equation}
\boldsymbol{F}_{Q}^{(i)} = \sum_{j \neq i}^{2} \frac{F_{Q} \boldsymbol{r}_{ij}}{r_{ij}} \bigg[ \frac{D^2}{r_{ij}^2} \bigg] \,,
\end{equation}
where $F_Q = Q^2 / (4 \pi \epsilon D^2)$ is the force scale of the electrostatic interactions.
To compute the optical forces $\boldsymbol{F}_{O}^{(i)}$, we apply the ray tracing approach \cite{Ashkin1992}.
The light source in our model is a divergent beam of power $P=$~\SI{1.0}{\watt} and half angle $\sigma =$~\SI{8.5e-3}{rad}.
The light intensity profile is Gaussian: $I(r) = P \, {\rm e}^{- r^2/(2 w^2)} / (2 \pi w^2)$, where $r$ is the distance to the centerline.
Setting the beam waist at $z = 0$, the beam width varies with $z$ as $w = \sigma z$.
We decompose this beam into several rays which reflect and refract at the air-glycerol interface following the Fresnel equations \cite{Ashkin1992}. By calculating the momentum change of the incoming versus outgoing rays, we obtain the optical forces on each droplet. Since the rays leaving the surface of droplet {\sf 1} may further strike the surface of droplet {\sf 2}, we also add their contributions to the net force of {\sf 2}, and vice versa. We integrate the droplet equations of motion in Eq.~(\ref{eq:eom}) to obtain the trajectories (Fig.~\ref{fig:SimulatedTrajectories} and \footnote{See Supplemental Material at \url{https://youtu.be/8xGxZVpvfN8} for a simulation of droplets juggling in the laser beam.}). By analyzing the droplet velocity spectra, we obtain an oscillation frequency of $32.5 \pm$\SI{2.0}{\hertz}. This corresponds to an orbital period of $30.8 \pm$\SI{1.1}{\milli\second} obtained from the velocity autocorrelations, in good agreement with the experimentally measured value.
%\begin{video}
%\href{https://youtu.be/8xGxZVpvfN8}{\includegraphics[width= 0.7\columnwidth, trim={40 0 40 10}, clip]{orbits_coulomb_ver_01.png}}
% \setfloatlink{https://youtu.be/8xGxZVpvfN8}%
% \caption{\label{vid:SimulatedTrajectories}%
%  Horizontal ($y$) and vertical ($z$) axes are distances measured from the beam waist at origin. Arrows indicate the instantaneous velocity of the particles. The color bar indicates time in milliseconds.
% }%
%\end{video}
\begin{figure}[htbp]
% trim the figure {left bottom right top}
\includegraphics[width= 0.6 \columnwidth, trim={40 0 40 10}, clip]{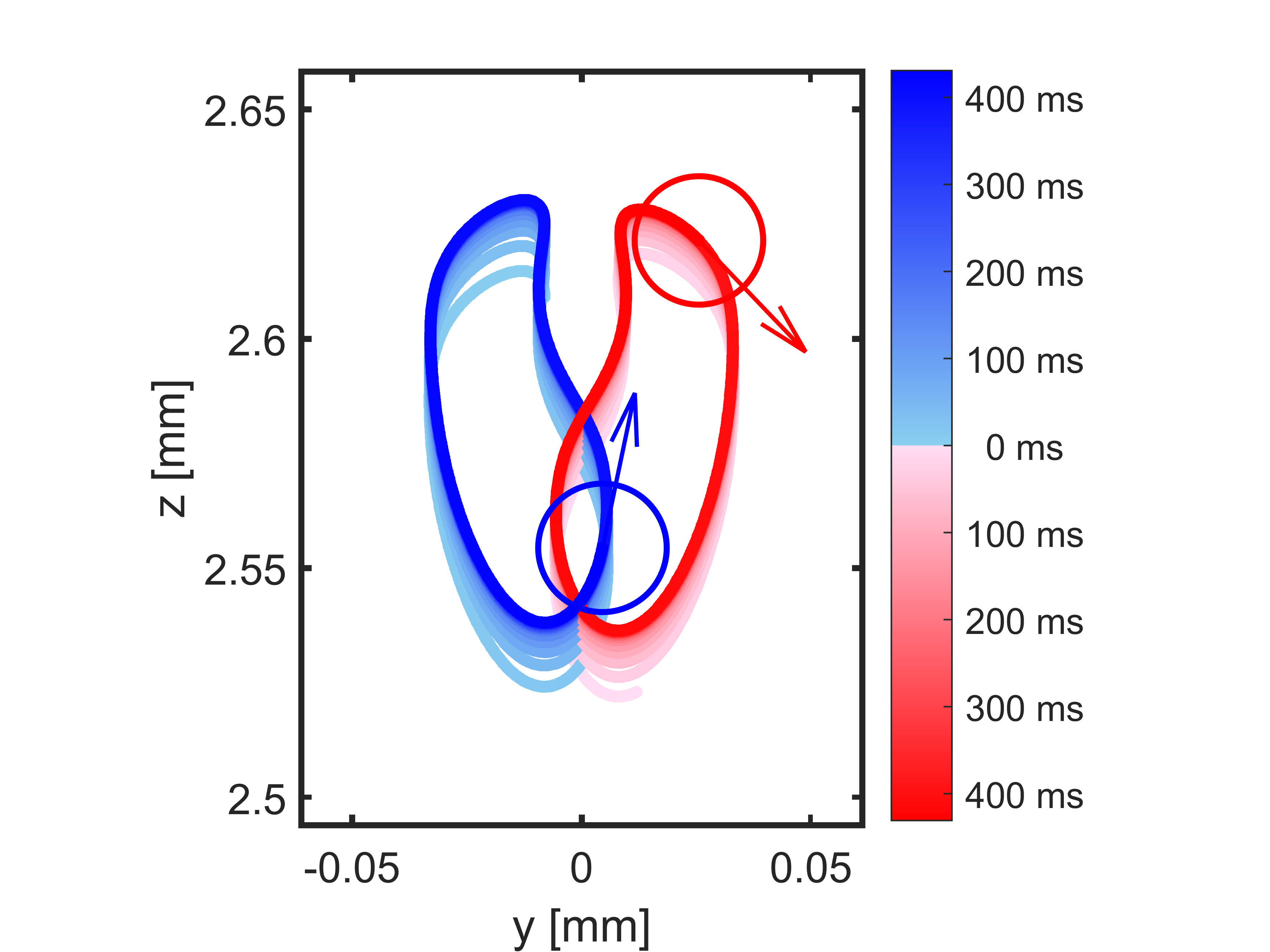}
\caption{\label{fig:SimulatedTrajectories} Horizontal ($y$) and vertical ($z$) axes are distances measured from the beam waist at origin. The beam polarization vector is perpendicular to the page. Arrows indicate the instantaneous velocity of the particles. The color bar indicates time in milliseconds.}
\end{figure}
\\
\indent
Having established that the model outlined above was in agreement with experiments, we were curious to test it against another effect we also observed. We had noticed that the droplets eventually settle in orbits in the same vertical plane that depends only on the direction of the polarization vector. In a separate experiment, we placed a half-wave plate in the path of the incident laser beam to rotate the polarization vector of light at a constant speed. When the half-wave plate continuously rotated through $45 \degree$ the polarization vector continuously rotated through $90 \degree$. We observed that the droplets continue to juggle while their plane of motion continuously rotates about the beam axis \footnote{See Supplemental Material at \url{https://youtu.be/XD4SzyT9itw} for a movie of the rotation of droplet orbits by rotating the polarization vector of light.}. The rotation ceased when the plane of motion lay perpendicular to the polarization vector. \\
\indent
Our model accounts in a natural way for the plane of motion. We illustrate the principal mechanism with Fig.~\ref{fig:PolarizationForces}.
We show two representative, linearly polarized rays propagating with equal power (Fig.~\ref{fig:PolarizationForces}a). They scatter at the interface of an off-center particle with different polarization direction.
The ray propagating from $P$ is $p$-polarized, whereas that from $S$ is $s$-polarized. The gradient force is directly proportional to the transmitted power \cite{Ashkin1992}, so $F_{\rm GRAD} \propto T^2$, where $T$ is the transmittance. Because the transmittance of $p$-polarized ray is greater than $s$-polarized ray (Fig.~\ref{fig:PolarizationForces}b), the net gradient force will restore the off-center particle to the centerline while pushing the particle towards the plane perpendicular to the polarization vector (which we term the $s$-plane) than towards the $p$-plane. This is the main mechanism which leads to the orbits settling in the $s$-plane. \\
\indent
To test the mechanism of alignment, we rotated the polarization vector of the incident light at a constant speed in our model. We observed that the particles gradually adjust their trajectories to lie on a plane that rotates with the polarization vector \footnote{See Supplemental Material at \url{https://youtu.be/43iK7zOQ4Qk} for a simulation of droplet motion in a beam with rotating polarization vector.}. To understand the stability of the alignment, we calculated the tangential component of the gradient force $F_T$ (that pushes perpendicular to the position vector $\boldsymbol{r}$) of a \SI{28}{\micro\metre} glycerol droplet as a function of the position vector $\boldsymbol{r}$ (Fig.~\ref{fig:PolarizationForces}c). The droplet is placed at a typical height of $z =$ \SI{2.6}{\milli\metre} from the beam waist. At any point on the $p$-plane (containing the wave vector $\boldsymbol{k}$ and the initial polarization vector $\boldsymbol{E}$) the tangential force diminishes. The droplet, however, is in unstable equilibrium. As soon as the droplet deviates from this plane, the tangential force pushes the droplet in the direction of $\boldsymbol{k} \times \boldsymbol{r} \, {\rm sgn} (\tan \theta)$, where $\theta$ is the polar angle of the position vector $\boldsymbol{r}$ from the $x$-axis.
When the droplet lies on the $s$-plane (containing the vectors $\boldsymbol{k}$ and $\boldsymbol{k} \times \boldsymbol{E}$) the tangential force again diminishes. On this plane, the droplet is in stable equilibrium. Therefore, if the droplet starts out in a position away from the center, the tangential force will always restore the droplet back to the $s$-plane. This restoring force in our ray-optics model is analogous to the alignment torque in the Rayleigh regime, in which particle sizes are much smaller than the wavelength of the incident light \cite{Haefner2009}. In the Rayleigh regime, Haefner et al. \cite{Haefner2009} analytically show that linearly polarized light can impart mechanical torque on a pair of particles and align their separation vector $\boldsymbol{r}$ perpendicular to the polarization vector $\boldsymbol{E}$. This curious alignment is also evident in our juggling droplets. \\
\indent
The inquisitive reader may well ask, ``Can we juggle small particles?'' We shall demonstrate this possibility using Moore \textit{et al.}'s pioneering experiment \cite{Moore2016}. To match their experimental conditions, consider silica particles of diameter $D = $~\SI{7}{\micro\metre}, density $\rho =$~\SI{2.65e3}{kg.m^{-3}} and index of refraction $n_2 = 1.45$ illuminated by a laser beam of power $P=$~\SI{0.5}{\watt}, wavelength $\lambda =$~\SI{1.5}{\micro\metre} and numerical aperture ${\rm NA}=0.1$ \cite{Note2}.
To estimate the electrostatic charge missing from their work, we assumed a constant surface charge density of \SI{1.5e-6}{C.m^{-2}} \cite{Nieh1988}, so that the net charge of a \SI{7}{\micro\metre} particle is $Q = $~\SI{2.3e-16}{\coulomb}. Under these conditions, the particles dance in complex patterns \footnote{See Supplemental Material at \url{https://youtu.be/mVgyN1Yibqs} for a simulation of the oscillations in Moore \textit{et al.}'s experiment.} resembling the lima\c{c}on trisectrix of D\"{u}rer and Pascal \footnote{See \url{http://mathworld.wolfram.com/Limacon.html}}.
As in juggling, the plane of motion lies perpendicular to the polarization vector. The motion in Moore \textit{et al.}'s experiment belongs to a regime where electrostatic forces contribute significantly to the dynamics and where ray optics is nearing its limits ($D \approx 5 \, \lambda$). In this regime, the particles play an anti-tug-of-war in which they attempt to push their way towards the center of the beam, but the strong electrostatic repulsion prevents them from getting there. We analysed the particle position spectra and obtained an oscillation frequency of \SI{2.6}{\hertz}, in close agreement with the \SI{3}{\hertz} frequency reported in \cite{Moore2016}.\\
\indent
Although the complete description of classical optics relies on solving the complex partial differential equations of electrodynamics, we have found in this work that the much simpler ray optics is sufficient at explaining most of the salient features. It is interesting to speculate the implications of our findings.
Might optical juggling be used for studying two-body interactions in juggling fountain \cite{Legere1998}, charge interactions in colloidal systems \cite{Brunner2004} and hydrodynamic interactions in two-dimensional systems \cite{Leonardo2008}? Our work demonstrates how well-studied physical systems can contain rich and undiscovered phenomena. A simple beam of light still holds an enduring fascination for us, so let there be light.

It is a pleasure to thank Jan Molacek and John Lawson for sharing their expertise, Raymond Shaw and Eberhard Bodenschatz for critical comments.
K. C. and D. H. are supported by the Knut and Alice Wallenberg Foundation Grant No. 2014.0048.
A. J. B. received support from the MaxSynBio Consortium, which is jointly funded by the Federal Ministry of Education and Research of Germany and the Max Planck Society.

%\bibliography{references}

\begin{thebibliography}{30}%
\makeatletter
\providecommand \@ifxundefined [1]{%
 \@ifx{#1\undefined}
}%
\providecommand \@ifnum [1]{%
 \ifnum #1\expandafter \@firstoftwo
 \else \expandafter \@secondoftwo
 \fi
}%
\providecommand \@ifx [1]{%
 \ifx #1\expandafter \@firstoftwo
 \else \expandafter \@secondoftwo
 \fi
}%
\providecommand \natexlab [1]{#1}%
\providecommand \enquote  [1]{``#1''}%
\providecommand \bibnamefont  [1]{#1}%
\providecommand \bibfnamefont [1]{#1}%
\providecommand \citenamefont [1]{#1}%
\providecommand \href@noop [0]{\@secondoftwo}%
\providecommand \href [0]{\begingroup \@sanitize@url \@href}%
\providecommand \@href[1]{\@@startlink{#1}\@@href}%
\providecommand \@@href[1]{\endgroup#1\@@endlink}%
\providecommand \@sanitize@url [0]{\catcode `\\12\catcode `\$12\catcode
  `\&12\catcode `\#12\catcode `\^12\catcode `\_12\catcode `\%12\relax}%
\providecommand \@@startlink[1]{}%
\providecommand \@@endlink[0]{}%
\providecommand \url  [0]{\begingroup\@sanitize@url \@url }%
\providecommand \@url [1]{\endgroup\@href {#1}{\urlprefix }}%
\providecommand \urlprefix  [0]{URL }%
\providecommand \Eprint [0]{\href }%
\providecommand \doibase [0]{http://dx.doi.org/}%
\providecommand \selectlanguage [0]{\@gobble}%
\providecommand \bibinfo  [0]{\@secondoftwo}%
\providecommand \bibfield  [0]{\@secondoftwo}%
\providecommand \translation [1]{[#1]}%
\providecommand \BibitemOpen [0]{}%
\providecommand \bibitemStop [0]{}%
\providecommand \bibitemNoStop [0]{.\EOS\space}%
\providecommand \EOS [0]{\spacefactor3000\relax}%
\providecommand \BibitemShut  [1]{\csname bibitem#1\endcsname}%
\let\auto@bib@innerbib\@empty
%</preamble>
\bibitem [{\citenamefont {Ashkin}(1970)}]{Ashkin1970}%
  \BibitemOpen
  \bibfield  {author} {\bibinfo {author} {\bibfnamefont {A.}~\bibnamefont
  {Ashkin}},\ }\href@noop {} {\bibfield  {journal} {\bibinfo  {journal} {Phys.
  Rev. Lett.}\ }\textbf {\bibinfo {volume} {24}},\ \bibinfo {pages} {156}
  (\bibinfo {year} {1970})}\BibitemShut {NoStop}%
\bibitem [{\citenamefont {Ashkin}\ and\ \citenamefont
  {Dziedzic}(1971)}]{Ashkin1971}%
  \BibitemOpen
  \bibfield  {author} {\bibinfo {author} {\bibfnamefont {A.}~\bibnamefont
  {Ashkin}}\ and\ \bibinfo {author} {\bibfnamefont {J.~M.}\ \bibnamefont
  {Dziedzic}},\ }\href@noop {} {\bibfield  {journal} {\bibinfo  {journal}
  {Appl. Phys. Lett.}\ }\textbf {\bibinfo {volume} {19}},\ \bibinfo {pages}
  {283} (\bibinfo {year} {1971})}\BibitemShut {NoStop}%
\bibitem [{\citenamefont {Ashkin}\ and\ \citenamefont
  {Dziedzic}(1974)}]{Ashkin1974}%
  \BibitemOpen
  \bibfield  {author} {\bibinfo {author} {\bibfnamefont {A.}~\bibnamefont
  {Ashkin}}\ and\ \bibinfo {author} {\bibfnamefont {J.~M.}\ \bibnamefont
  {Dziedzic}},\ }\href@noop {} {\bibfield  {journal} {\bibinfo  {journal}
  {Appl. Phys. Lett.}\ }\textbf {\bibinfo {volume} {24}},\ \bibinfo {pages}
  {586} (\bibinfo {year} {1974})}\BibitemShut {NoStop}%
\bibitem [{\citenamefont {Ashkin}(1975)}]{Ashkin1975}%
  \BibitemOpen
  \bibfield  {author} {\bibinfo {author} {\bibfnamefont {A.}~\bibnamefont
  {Ashkin}},\ }\href@noop {} {\bibfield  {journal} {\bibinfo  {journal}
  {Science}\ }\textbf {\bibinfo {volume} {187}},\ \bibinfo {pages} {1073}
  (\bibinfo {year} {1975})}\BibitemShut {NoStop}%
\bibitem [{\citenamefont {Ashkin}\ \emph {et~al.}(1986)\citenamefont {Ashkin},
  \citenamefont {Dziedzic}, \citenamefont {Bjorkholm},\ and\ \citenamefont
  {Chu}}]{Ashkin1986}%
  \BibitemOpen
  \bibfield  {author} {\bibinfo {author} {\bibfnamefont {A.}~\bibnamefont
  {Ashkin}}, \bibinfo {author} {\bibfnamefont {J.~M.}\ \bibnamefont
  {Dziedzic}}, \bibinfo {author} {\bibfnamefont {J.~E.}\ \bibnamefont
  {Bjorkholm}}, \ and\ \bibinfo {author} {\bibfnamefont {S.}~\bibnamefont
  {Chu}},\ }\href@noop {} {\bibfield  {journal} {\bibinfo  {journal} {Opt.
  Lett.}\ }\textbf {\bibinfo {volume} {11}},\ \bibinfo {pages} {288} (\bibinfo
  {year} {1986})}\BibitemShut {NoStop}%
\bibitem [{\citenamefont {Ashkin}(1992)}]{Ashkin1992}%
  \BibitemOpen
  \bibfield  {author} {\bibinfo {author} {\bibfnamefont {A.}~\bibnamefont
  {Ashkin}},\ }\href@noop {} {\bibfield  {journal} {\bibinfo  {journal}
  {Biophys. J.}\ }\textbf {\bibinfo {volume} {61}},\ \bibinfo {pages} {569}
  (\bibinfo {year} {1992})}\BibitemShut {NoStop}%
\bibitem [{\citenamefont {Carmona-Sosa}\ and\ \citenamefont
  {Quinto-Su}(2016)}]{CarmonaSosa2016}%
  \BibitemOpen
  \bibfield  {author} {\bibinfo {author} {\bibfnamefont {V.}~\bibnamefont
  {Carmona-Sosa}}\ and\ \bibinfo {author} {\bibfnamefont {P.~A.}\ \bibnamefont
  {Quinto-Su}},\ }\href@noop {} {\bibfield  {journal} {\bibinfo  {journal} {J.
  Opt.}\ }\textbf {\bibinfo {volume} {18}},\ \bibinfo {pages} {105301}
  (\bibinfo {year} {2016})}\BibitemShut {NoStop}%
\bibitem [{\citenamefont {Moore}\ \emph {et~al.}(2016)\citenamefont {Moore},
  \citenamefont {Martin}, \citenamefont {Maayani}, \citenamefont {Kim},
  \citenamefont {Chandrahalim}, \citenamefont {Eichenfield}, \citenamefont
  {Martin},\ and\ \citenamefont {Carmon}}]{Moore2016}%
  \BibitemOpen
  \bibfield  {author} {\bibinfo {author} {\bibfnamefont {J.}~\bibnamefont
  {Moore}}, \bibinfo {author} {\bibfnamefont {L.~L.}\ \bibnamefont {Martin}},
  \bibinfo {author} {\bibfnamefont {S.}~\bibnamefont {Maayani}}, \bibinfo
  {author} {\bibfnamefont {K.~H.}\ \bibnamefont {Kim}}, \bibinfo {author}
  {\bibfnamefont {H.}~\bibnamefont {Chandrahalim}}, \bibinfo {author}
  {\bibfnamefont {M.}~\bibnamefont {Eichenfield}}, \bibinfo {author}
  {\bibfnamefont {I.~R.}\ \bibnamefont {Martin}}, \ and\ \bibinfo {author}
  {\bibfnamefont {T.}~\bibnamefont {Carmon}},\ }\href@noop {} {\bibfield
  {journal} {\bibinfo  {journal} {Opt. Express}\ }\textbf {\bibinfo {volume}
  {24}},\ \bibinfo {pages} {2850} (\bibinfo {year} {2016})}\BibitemShut
  {NoStop}%
\bibitem [{\citenamefont {Mitra}\ \emph {et~al.}(2018)\citenamefont {Mitra},
  \citenamefont {Brown}, \citenamefont {Bernot}, \citenamefont {Defrances},\
  and\ \citenamefont {Talghader}}]{Mitra2018}%
  \BibitemOpen
  \bibfield  {author} {\bibinfo {author} {\bibfnamefont {T.}~\bibnamefont
  {Mitra}}, \bibinfo {author} {\bibfnamefont {A.~K.}\ \bibnamefont {Brown}},
  \bibinfo {author} {\bibfnamefont {D.~M.}\ \bibnamefont {Bernot}}, \bibinfo
  {author} {\bibfnamefont {S.}~\bibnamefont {Defrances}}, \ and\ \bibinfo
  {author} {\bibfnamefont {J.~J.}\ \bibnamefont {Talghader}},\ }\href@noop {}
  {\bibfield  {journal} {\bibinfo  {journal} {Opt. Express}\ }\textbf {\bibinfo
  {volume} {26}},\ \bibinfo {pages} {6639} (\bibinfo {year}
  {2018})}\BibitemShut {NoStop}%
\bibitem [{Note1()}]{Note1}%
  \BibitemOpen
  \bibinfo {note} {See Supplemental Material at \protect \url
  {https://youtu.be/ZyXBM8B0Md0} for a movie of the droplets juggling in the
  laser beam. It has come to our attention that Moore \protect \textit {et al.}
  observed an oscillations under quite different conditions. Our theory not
  only explains their observation, but also makes predictions on their particle
  charge and oscillation frequency.}\BibitemShut {Stop}%
\bibitem [{\citenamefont {Polster}(2003)}]{Polster2003}%
  \BibitemOpen
  \bibfield  {author} {\bibinfo {author} {\bibfnamefont {B.}~\bibnamefont
  {Polster}},\ }\href@noop {} {\emph {\bibinfo {title} {{The Mathematics of
  Juggling}}}}\ (\bibinfo  {publisher} {Springer-Verlag},\ \bibinfo {address}
  {New York},\ \bibinfo {year} {2003})\BibitemShut {NoStop}%
\bibitem [{Note2()}]{Note2}%
  \BibitemOpen
  \bibinfo {note} {See Supplemental Material at [URL] for additional details,
  which includes Refs.~\cite {Atherton1999, Mordant2004,
  Shampine1997}.}\BibitemShut {Stop}%
\bibitem [{\citenamefont {Bailey}(1988)}]{Bailey1988}%
  \BibitemOpen
  \bibfield  {author} {\bibinfo {author} {\bibfnamefont {A.~G.}\ \bibnamefont
  {Bailey}},\ }\href@noop {} {\emph {\bibinfo {title} {{Electrostatic Spraying
  of Liquids}}}}\ (\bibinfo  {publisher} {Wiley},\ \bibinfo {address} {New
  York},\ \bibinfo {year} {1988})\BibitemShut {NoStop}%
\bibitem [{Note3()}]{Note3}%
  \BibitemOpen
  \bibinfo {note} {See Supplemental Material at \protect \url
  {https://youtu.be/UiqqxNjS_v8} for a movie of the droplets coalescing in the
  laser beam.}\BibitemShut {Stop}%
\bibitem [{Note4()}]{Note4}%
  \BibitemOpen
  \bibinfo {note} {See Supplemental Material at \protect \url
  {https://youtu.be/uZPhUhgpxxk} for a movie of the formation of juggling
  movement.}\BibitemShut {Stop}%
\bibitem [{Note5()}]{Note5}%
  \BibitemOpen
  \bibinfo {note} {We allude to the elegant quote by John A. Wheeler,
  ``Spacetime tells matter how to move; matter tells spacetime how to
  curve''}\BibitemShut {NoStop}%
\bibitem [{\citenamefont {Briels}(1994)}]{Briels1994}%
  \BibitemOpen
  \bibfield  {author} {\bibinfo {author} {\bibfnamefont {W.}~\bibnamefont
  {Briels}},\ }\href@noop {} {\emph {\bibinfo {title} {{Theory of Polymer
  Dynamics}}}}\ (\bibinfo  {publisher} {Uppsala University},\ \bibinfo
  {address} {Uppsala},\ \bibinfo {year} {1994})\BibitemShut {NoStop}%
\bibitem [{Note6()}]{Note6}%
  \BibitemOpen
  \bibinfo {note} {See Supplemental Material at \protect \url
  {https://youtu.be/8xGxZVpvfN8} for a simulation of droplets juggling in the
  laser beam.}\BibitemShut {Stop}%
\bibitem [{Note7()}]{Note7}%
  \BibitemOpen
  \bibinfo {note} {See Supplemental Material at \protect \url
  {https://youtu.be/XD4SzyT9itw} for a movie of the rotation of droplet orbits
  by rotating the polarization vector of light.}\BibitemShut {Stop}%
\bibitem [{Note8()}]{Note8}%
  \BibitemOpen
  \bibinfo {note} {See Supplemental Material at \protect \url
  {https://youtu.be/43iK7zOQ4Qk} for a simulation of droplet motion in a beam
  with rotating polarization vector.}\BibitemShut {Stop}%
\bibitem [{\citenamefont {Haefner}\ \emph {et~al.}(2009)\citenamefont
  {Haefner}, \citenamefont {Sukhov},\ and\ \citenamefont
  {Dogariu}}]{Haefner2009}%
  \BibitemOpen
  \bibfield  {author} {\bibinfo {author} {\bibfnamefont {D.}~\bibnamefont
  {Haefner}}, \bibinfo {author} {\bibfnamefont {S.}~\bibnamefont {Sukhov}}, \
  and\ \bibinfo {author} {\bibfnamefont {A.}~\bibnamefont {Dogariu}},\
  }\href@noop {} {\bibfield  {journal} {\bibinfo  {journal} {Phys. Rev. Lett.}\
  }\textbf {\bibinfo {volume} {103}},\ \bibinfo {pages} {173602} (\bibinfo
  {year} {2009})}\BibitemShut {NoStop}%
\bibitem [{\citenamefont {Nieh}\ and\ \citenamefont {Nguyen}(1988)}]{Nieh1988}%
  \BibitemOpen
  \bibfield  {author} {\bibinfo {author} {\bibfnamefont {S.}~\bibnamefont
  {Nieh}}\ and\ \bibinfo {author} {\bibfnamefont {T.}~\bibnamefont {Nguyen}},\
  }\href@noop {} {\bibfield  {journal} {\bibinfo  {journal} {J. Electrostat.}\
  }\textbf {\bibinfo {volume} {21}},\ \bibinfo {pages} {99} (\bibinfo {year}
  {1988})}\BibitemShut {NoStop}%
\bibitem [{Note9()}]{Note9}%
  \BibitemOpen
  \bibinfo {note} {See Supplemental Material at \protect \url
  {https://youtu.be/mVgyN1Yibqs} for a simulation of the oscillations in Moore
  \protect \textit {et al.}'s experiment.}\BibitemShut {Stop}%
\bibitem [{Note10()}]{Note10}%
  \BibitemOpen
  \bibinfo {note} {See \protect \url
  {http://mathworld.wolfram.com/Limacon.html}}\BibitemShut {NoStop}%
\bibitem [{\citenamefont {Legere}\ and\ \citenamefont
  {Gibble}(1998)}]{Legere1998}%
  \BibitemOpen
  \bibfield  {author} {\bibinfo {author} {\bibfnamefont {R.}~\bibnamefont
  {Legere}}\ and\ \bibinfo {author} {\bibfnamefont {K.}~\bibnamefont
  {Gibble}},\ }\href@noop {} {\bibfield  {journal} {\bibinfo  {journal} {Phys.
  Rev. Lett.}\ }\textbf {\bibinfo {volume} {81}},\ \bibinfo {pages} {5780}
  (\bibinfo {year} {1998})}\BibitemShut {NoStop}%
\bibitem [{\citenamefont {Brunner}\ \emph {et~al.}(2004)\citenamefont
  {Brunner}, \citenamefont {Dobnikar}, \citenamefont {von Gr{\"{u}}nberg},\
  and\ \citenamefont {Bechinger}}]{Brunner2004}%
  \BibitemOpen
  \bibfield  {author} {\bibinfo {author} {\bibfnamefont {M.}~\bibnamefont
  {Brunner}}, \bibinfo {author} {\bibfnamefont {J.}~\bibnamefont {Dobnikar}},
  \bibinfo {author} {\bibfnamefont {H.-H.}\ \bibnamefont {von Gr{\"{u}}nberg}},
  \ and\ \bibinfo {author} {\bibfnamefont {C.}~\bibnamefont {Bechinger}},\
  }\href@noop {} {\bibfield  {journal} {\bibinfo  {journal} {Phys. Rev. Lett.}\
  }\textbf {\bibinfo {volume} {92}},\ \bibinfo {pages} {078301} (\bibinfo
  {year} {2004})}\BibitemShut {NoStop}%
\bibitem [{\citenamefont {Leonardo}\ \emph {et~al.}(2008)\citenamefont
  {Leonardo}, \citenamefont {Keen}, \citenamefont {Ianni}, \citenamefont
  {Leach}, \citenamefont {Padgett},\ and\ \citenamefont
  {Ruocco}}]{Leonardo2008}%
  \BibitemOpen
  \bibfield  {author} {\bibinfo {author} {\bibfnamefont {R.~D.}\ \bibnamefont
  {Leonardo}}, \bibinfo {author} {\bibfnamefont {S.}~\bibnamefont {Keen}},
  \bibinfo {author} {\bibfnamefont {F.}~\bibnamefont {Ianni}}, \bibinfo
  {author} {\bibfnamefont {J.}~\bibnamefont {Leach}}, \bibinfo {author}
  {\bibfnamefont {M.~J.}\ \bibnamefont {Padgett}}, \ and\ \bibinfo {author}
  {\bibfnamefont {G.}~\bibnamefont {Ruocco}},\ }\href@noop {} {\bibfield
  {journal} {\bibinfo  {journal} {Phys. Rev. E}\ }\textbf {\bibinfo {volume}
  {78}},\ \bibinfo {pages} {031406} (\bibinfo {year} {2008})}\BibitemShut
  {NoStop}%
\bibitem [{\citenamefont {Atherton}\ and\ \citenamefont
  {Kerbyson}(1999)}]{Atherton1999}%
  \BibitemOpen
  \bibfield  {author} {\bibinfo {author} {\bibfnamefont {T.~J.}\ \bibnamefont
  {Atherton}}\ and\ \bibinfo {author} {\bibfnamefont {D.~J.}\ \bibnamefont
  {Kerbyson}},\ }\href@noop {} {\bibfield  {journal} {\bibinfo  {journal}
  {Image Vis. Comput.}\ }\textbf {\bibinfo {volume} {17}},\ \bibinfo {pages}
  {795} (\bibinfo {year} {1999})}\BibitemShut {NoStop}%
\bibitem [{\citenamefont {Mordant}\ \emph {et~al.}(2004)\citenamefont
  {Mordant}, \citenamefont {Crawford},\ and\ \citenamefont
  {Bodenschatz}}]{Mordant2004}%
  \BibitemOpen
  \bibfield  {author} {\bibinfo {author} {\bibfnamefont {N.}~\bibnamefont
  {Mordant}}, \bibinfo {author} {\bibfnamefont {A.~M.}\ \bibnamefont
  {Crawford}}, \ and\ \bibinfo {author} {\bibfnamefont {E.}~\bibnamefont
  {Bodenschatz}},\ }\href@noop {} {\bibfield  {journal} {\bibinfo  {journal}
  {Physica D}\ }\textbf {\bibinfo {volume} {193}},\ \bibinfo {pages} {245}
  (\bibinfo {year} {2004})}\BibitemShut {NoStop}%
\bibitem [{\citenamefont {Shampine}\ and\ \citenamefont
  {Reichelt}(1997)}]{Shampine1997}%
  \BibitemOpen
  \bibfield  {author} {\bibinfo {author} {\bibfnamefont {L.~F.}\ \bibnamefont
  {Shampine}}\ and\ \bibinfo {author} {\bibfnamefont {M.~W.}\ \bibnamefont
  {Reichelt}},\ }\href@noop {} {\bibfield  {journal} {\bibinfo  {journal} {SIAM
  J. Sci. Comput.}\ }\textbf {\bibinfo {volume} {18}},\ \bibinfo {pages} {1}
  (\bibinfo {year} {1997})}\BibitemShut {NoStop}%
\end{thebibliography}

\end{document}